# Decoherence of Single-Excitation Entanglement over DLCZ Quantum Networks Caused by Slow-Magnetic-Field Fluctuations and Protection Approach

*Can Sun, Ya Li, Yi-bo Hou, Min-jie Wang, Shu-jing Li\*, Hai Wang\**

C. Sun, Y. Li, Y. Hou, M. Wang, S. Li. and H. Wang

1. State Key Laboratory of Quantum Optics and Quantum Optics Devices, Institute of Opto-Electronics. 2.Collaborative Innovation Center of Extreme Optics, Shanxi University, Taiyuan 030006, China

*Corresponding author email: lishujing@sxu.edu.cn and wanghai@sxu.edu.cn

**Data availability**

All the data and calculations that support the findings of this study are available from the corresponding author upon reasonable request.

**Abstract**

Atomic spin-wave (SW) memory is a building block for quantum repeater. However, due to decoherence caused by atomic motions and magnetic-field gradients, SW retrieval efficiency decays with storage time. Slow magnetic-field fluctuations (SMFFs) lead to shot-to-shot phase noises of SWs, but have not been considered in previous repeater protocols. Here, we develop a SW model including a homogeneous phase noise caused by SMFFs, and then reveal that such phase noise induces decoherence of single-excitation entanglement between two atomic ensembles in a repeater link. For verify this conclusion, we experimentally prepare single-excitation entanglement between two SWs in a cold atomic ensemble and observe entanglement decoherence induced by SMFFs. Limited by SMFFs, the evaluated lifetime of the single-excitation entanglement over a repeater link is ~135 ms, even if the link uses optical-lattice atoms as nodes. We then present an approach that may overcome such limitation and extend this lifetime to ~1.7 s.

1. Introduction

Quantum repeaters (QRs) hold promise for realizing long-distance quantum communication [1-2], large-scale quantum networks (QNs) [1-6], and longer-baseline telescopes [7]. In QR protocol [8], a long distance, over which entanglement is required to be distributed, is divided into N shorter elementary links. Entanglement is at first created between two quantum memories (nodes) in elementary links and then extended over full distance via entanglement swapping. In the past two decades, various physical systems [9] such as atomic ensembles [2,10] and single quantum systems, including single atoms [11-12], ions [13-15] and solid-state spins [16-18] have been developed as memory nodes. The atomic-ensemble-based quantum memories (QMs) can be efficiently mapped into photons [1-3,19] and then hold promise for entanglement swapping. The well-known Duan-Lukin-Cirac-Zoller (DLCZ) protocol is a QR architecture based on atomic ensembles [1-2]. It creates spin-wave (SW) memory via spontaneous Raman emission, which simultaneously emits a Stoke photon correlated with the SW memory. Significant progress has been achieved on experimental generations of non-classically correlated or entangled pairs of a photon and a SW via DLCZ protocol [20-41]. Alternately, the correlated or entangled pairs can be achieved by using SPDC process together with quantum memories [42-45], where, the quantum memories are realized via "absorptive" ("read-write" [46]) storage schemes that are based on electromagnetically induced transparency, atomic frequency comb and off-resonant cascaded absorption *et. al*. With two spin-wave-photon pairs, single-excitation entanglement states between two memories (nodes), which form an elementary link, are generated [47-49] via single-photon interferometer (SPI).

Entanglement between two memories can also be established by two-photon-interference (TPI) scheme [49-51], which uses atom-photon entanglement in each node and removes the requirement for the long-distance phase stability in SPI schemes [52]. However, TPI scheme requires detecting two photons, which leads to a much low probability in elementary entanglement generation compared to that via SPI [50].

Entanglement between two remote memories in a link has been demonstrated with atomic ensembles [47-51,53-56] or single quantum systems including individual atoms [57-58], ions [15] and solid-state spins [16-18]. Also, by distributing entangled pairs of two photons [59] or of a SW and a photon over two nodes as links [60], entanglement connections between repeater links have been experimentally demonstrated. So far, QRs still face the challenges including the deterministic entanglement generation between two atomic memories with a separation long distance (for example 100 km) and the connection of two or more repeater links formed by matter nodes. To overcome such challenges, it is required to store entanglement for a long time and effectively retrieve the photons stored in memories [3].

To achieve long-lived QMs based on atomic ensemble, decoherence mechanisms of SWs in cold atomic ensembles have been studied, and two main decoherence processes have been revealed [22-24]. One process is due to atomic motions, which make atoms moving out spin-wave spatial mode in the atomic ensembles or lead to dephasing of spin wave [24]. Another process results from gradients of magnetic fields [22-23], which make spin transitions of the atoms at different positions experience inhomogeneous broadening and then lead to a loss of quantum phase in collective superpositions. Atomic-motion-induced decoherence can be suppressed either by lengthening spin-wave wavelengths [24,39,41,61] or confining the atoms in optical lattices [26-27,36,38]. Inhomogeneous-broadening-induced decoherence can be suppressed by storing SWs in magnetic-field-insensitive (or "clock") spin transitions [23,27,36,38,39,41,60]. Based on these schemes of suppressing decoherence, the storage times at $1/e$ retrieval efficiency for spin-wave QMs reach to ~400 ms in optical-lattice atoms [38]. It has been experimentally demonstrated that fidelities of entanglement between two ensembles [25] and polarization atom-photon entanglement [35] decay with time, respectively, due to the decoherence processes.

On the other hand, slow magnetic-field fluctuations (SMFFs) will induce shot-to-shot phase fluctuations of an atomic memory. In quantum systems such as trapped single atoms and NV centers, the phase noise of the atomic memory qubit caused by SMFFs has been pointed out [11,12,17], and dynamical decoupling strategies have been developed to suppress such phase noise [12,17]. Quantum storages of weak coherence light with arbitrary polarization based on EIT scheme in ultra-cold atomic ensemble (BEC) have been demonstrated and the polarization fidelity decay caused by SMFFs has been reported [63]. By compensating the magnetic-field fluctuations with an open-loop feed-forward circuit, the long-lived optical polarization storage was achieved in that experiment. In a recent experiment on entanglement generation between two atomic ensembles [56], decoherence of a SW memory for polarization qubit caused by SMFFs is observed, where, SMFFs are mainly produced by ac noise of Helmholtz coils. By synchronizing experimental sequence with the main electricity frequency 50-Hz, the qubit decoherence was significantly suppressed in that experiment. However, so far, the SW phase noise caused by SMFFs has not been modeled and its influence on entanglement between two atomic ensembles in DLCZ-like links has not been demonstrated.

In the presented work, we develop a SW model that includes phase noise induced by SMFFs. Such phase noise is homogeneous for each SW supposition term and then will not affect SW retrieval efficiency at any storage time. While, SMFFs will lead to phase-noise difference between two ensembles in a repeater link and then degrade the single-excitation entanglement between the ensembles. For verifying this conclusion, we experimentally generate two single-excitation entanglement states between SWs in a cold atomic ensemble. By measuring the visibility of interference between entangled two SWs, where, two SWs are stored in a magnetic-field-sensitive and a "clock" spin transition, respectively, we observe that coherence between the SWs decays with storage time. We also demonstrate that the phase-noise difference between entangled two SWs is eliminated when both SWs are stored in the same magnetic-field-sensitive transitions. Based on our SW model and previous experimental data in Refs. [38, 62], we predict that the lifetime of single-excitation entanglement between two ensembles is mainly limited to decoherence induced by SMFFs. Even if the link uses optical-lattice atoms as memory nodes, the lifetime of entanglement over a repeater link is limited by ~135 ms. We finally propose a scheme that can cancel the phase-noise difference, where, a single DC supply is used to drive the Helmholtz coils of two nodes and then make the magnetic fields fluctuation at both nodes be the same.

## 2. Theoretical analysis: decoherence of single-excitation entanglement between two atomic ensembles caused by SMFFs

At first, we develop a SW model that includes amplitude and phase decoherence. As shown in **Figure 1**, the generation of SW is based on DLCZ process. A write laser pulse is applied to an ensemble of $N$ atoms with three-level structure, which creates a non-classically correlated pairs of a Stokes photon and a spin wave. When excitation probability $\chi \ll 1$, the

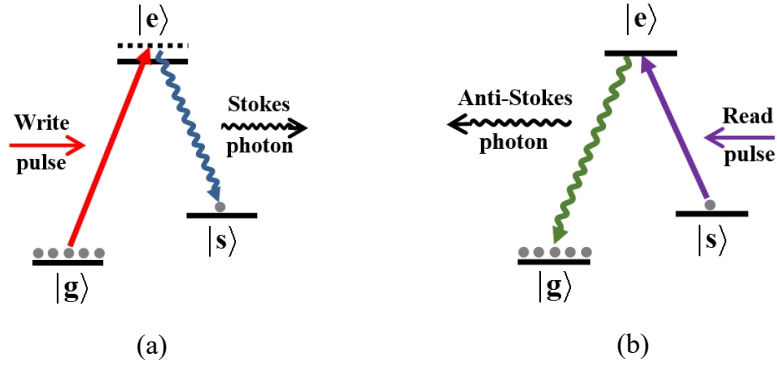

(a)　　　　　　　　　　　　(b)

**Figure 1.** Atomic levels related to the write (a) and read (b) processes in DLCZ scheme, respectively.

atom-photon system is written as [2,25]

$$\Phi_{a,p} = |0\rangle + \sqrt{\chi}|1\rangle_p |\psi_a(0)\rangle + O(\chi) \tag{1}$$

where, $|0\rangle$ denotes the vacuum part, $|1\rangle_p$ one photon in the Stokes field, $|\psi_a(0)\rangle$ represents a SW formed by a coherent superposition of all the possible states with $N$-1 atoms in $|g\rangle$ and one atom in $|s\rangle$ and is written as,

$$|\psi_a(0)\rangle = \frac{1}{\sqrt{N}} \sum_{j=1}^{N} e^{-i\Delta k z_j(0)} |g\rangle_1 |g\rangle_2 \cdots |s\rangle_j \cdots |g\rangle_N \tag{2}$$

where, $z_j(0)$ is the $j$-th atom coordinate at storage time $t$=0, $\Delta k = (k_w - k_s)$ the SW wave-vector, $k_w$ is the wave-vector of the write beam, whose frequency is slightly detuned from the $|g\rangle \to |e\rangle$ transition, $k_s$ the wave-vector of the Stokes photon, which is emitted on the $|g\rangle \to |e\rangle$ transition. The magnetic field in the atoms is generated by applying a DC supply to Helmholtz coils. It is along $z$ direction and can be written as $\vec{B} = (B_0 + B'z + \delta B(t,z))\vec{z}$, where, $B_0$ is the bias magnetic field, $B'$ is the gradient of the magnetic field, $\delta B(t,z)$ represents slow magnetic-field fluctuations (SMFFs). After storage time $t$, the spin wave is evolved into

$$|\psi_a(t)\rangle = \frac{1}{\sqrt{N}} e^{-i\omega_0 t} \sum_{j=1}^{N} e^{-i\Delta\omega(z_j)t - i\Delta k\left[z_j(0) + \delta z_j(t)\right] - i\delta\varphi(z_j,t)} |g\rangle_1 |g\rangle_2 \cdots |s\rangle_j \cdots |g\rangle_N \quad (3)$$

where, $\omega_0$ is the Larmor frequency of the used spin transition $|g, m_{Fg}\rangle \leftrightarrow |s, m_{Fs}\rangle$ caused by the bias field $B_0$, $\Delta\omega(z_j) = 2\pi\mu' B' z_j$ denotes Larmor frequency shifts of the atoms at the position $z_j$, which results in the inhomogeneous broadening of the spin transition, $\mu' = d\nu/dB$ corresponds to the sensitivity of the spin transition $|g, m_{Fg}\rangle \leftrightarrow |s, m_{Fs}\rangle$ to the magnetic field, $\nu$ is the frequency of the spin transition, and $\delta\varphi(z_j,t) = 2\pi \int_0^t \mu' \delta B(z_j, t')dt'$ denotes the phase noise of the atom induced by SMFF $\delta B(z_j,t)$ at the position $z_j$. $\delta B(z_j,t)$ mainly results from the current fluctuation of the DC supply and also includes a small ac noise of environment field. The frequencies of the coils range from 200 Hz to 0 Hz [63], and the wavelength of the magnetic-field fluctuations $\delta B(t,z)$ is more than $1500\ km$, which far exceeds the length (~5mm) of the ensemble. So, SMFF remains the same value for all of the atoms, i.e., $\delta B(z_j,t) = \delta B(t)$ and $\delta\varphi(z_j,t) = \delta\varphi(t) = 2\pi \int_0^t \mu' \delta B(t')dt'$. Thus, the SW is rewritten as

$$|\psi_a(t)\rangle = \frac{1}{\sqrt{N}} e^{-i\omega_0 t - i\delta\varphi(t)} \sum_{j=1}^{N} e^{-i\Delta\omega(z_j)t - i\Delta k\left[z_j(0) + \delta z_j(t)\right]} |g\rangle_1 |g\rangle_2 \cdots |s\rangle_j \cdots |g\rangle_N. \quad (4)$$

As pointed out in Ref. [64], the main contributions of the phase noise $\delta\varphi(t)$ can be assumed to result from shot-to-shot fluctuations $\delta B(t)$, i.e., for an individual trial with a storage time $t$, $\delta\varphi(t) = 2\pi\mu' \delta B t$. We assume that the magnetic-field fluctuations $\delta B$ follow Lorentzian distribution:

$$P(\delta B) = \frac{\sigma_B / \pi}{\delta B^2 + \sigma_B^2},\tag{5}$$

where, $\sigma_B$ is the rms width of the Lorentzian distribution. The spin wave at storage time $t$ can be rewritten as:

$$|\psi_a(t)\rangle = |\psi_a(0)\rangle\langle\psi_a(0)|\psi_a(t)\rangle = e^{-i\omega_0 t - i\delta\varphi(t)} D(t)|\psi_a(0)\rangle,\tag{6}$$

where, $D(t) = \sum_{j=1}^{N} e^{-i\Delta\omega(z_j)t - i\Delta k v_j t} = e^{-t^2/2\tau_1^2} e^{-t^2/2\tau_2^2} = e^{-t^2/2\tau_D^2}$, which denotes SW amplitude factor, where, $\tau_1$ and $\tau_2$ are the lifetimes limited by decoherence arising from atomic random motions (see Ref.[24] or Supporting information) and the inhomogeneous broadening of the spin transition caused by the gradient $B'$ (see Ref.[23] or Supporting information), respectively, $\tau_D = \left(\frac{\tau_1 \tau_2}{\sqrt{\tau_1^2 + \tau_2^2}}\right)$ is the total lifetime. The SW can be retrieved by a read pulse with the retrieval efficiency [24]

$$\gamma(t) = \gamma_0 |\langle\psi_a(0)|\psi_a(t)\rangle|^2 = \gamma_0 |D(t)|^2 = \gamma_0 e^{-t^2/\tau_D^2},\tag{7}$$

where, $\gamma_0 \leq 1$ denotes the zero-delay retrieval efficiency. The Eq. (6) shows that SMFFs have not led to decay of SW retrieval efficiency. However, we identify that it will induce decoherence of the single-excitation entangled state between two or multiple SW memories.

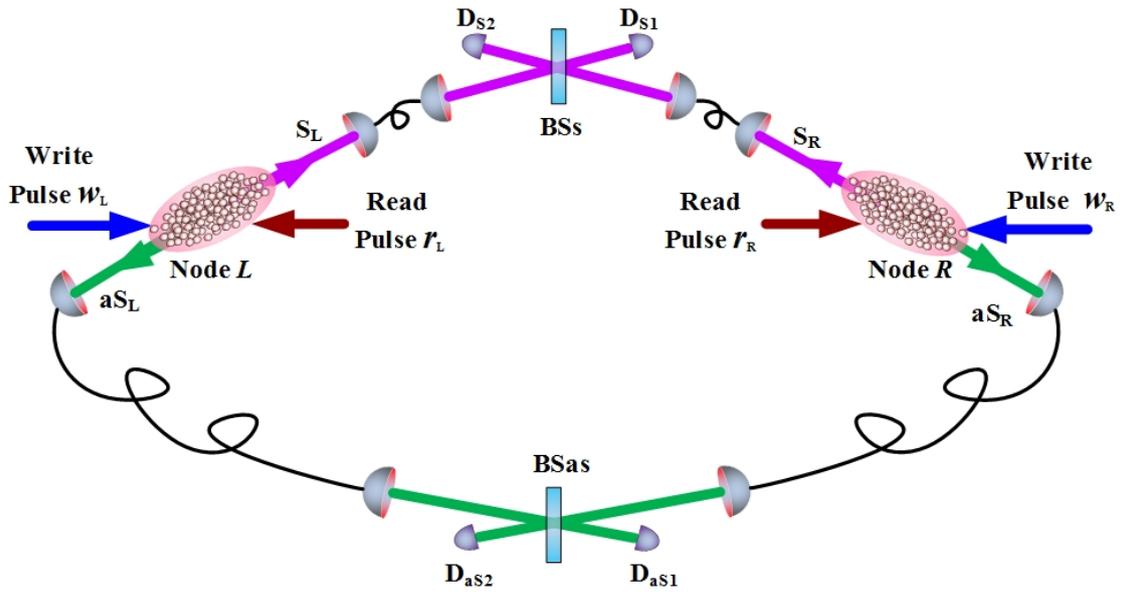

**Figure 2.** Single-excitation entanglement generation between two atomic ensembles *R* and *L* in a repeater link. A write (read) laser pulse is split into write (read) pulses $w_L$ ($r_L$) and $w_R$ ($r_R$) by a beam splitter BS$_w$ (BS$_r$). For simplicity, we didn't mark BS$_w$ (BS$_r$) in the figure. The write and read pulses $w_R$ and $r_R$ ($w_L$ and $r_L$) counter propagate and are applied to the ensembles *R* (*L*). The write pulse $w_R$ ($w_L$) creates the Stokes field $a_{S_R}$ ($a_{S_L}$) and SW $|1_R\rangle$ ($|1_L\rangle$) respectively. The Stokes field $a_{S_R}$ and $a_{S_L}$ are sent to a 50/50 beam splitter (BS$_S$). The SWs $|1_R\rangle$ and $|1_L\rangle$ are retrieved by read pulses $r_R$ and $r_L$, respectively. The retrieved fields $a_{aS_R}$ and $a_{aS_L}$ are sent to another 50/50 beam splitter (BS$_{aS}$). The beam splitters BS$_S$ and BS$_{aS}$ (BS$_W$ and BS$_R$) form Stokes-anti-Stokes field (write-read laser) interferometer. The difference of two arms of each interferometer is actively stabilized. Single-photon detectors D$_{S1}$ and D$_{S2}$ (D$_{aS1}$ and D$_{aS2}$) are used to detect the Stokes (anti-Stokes) fields.

**Figure 2** shows the sketch of generating single-excitation entanglement between two remote atomic ensembles via DLCZ protocol. Two ensembles *R* and *L*, which use the same-species atoms, are simultaneously applied by write pulses $w_R$ and $w_L$, respectively. The emitted Stokes fields $a_{S_R}$ and $a_{S_L}$ from the ensembles *R* and *L* are sent to a center station and combined at a 50/50 beam splitter (labeled as BS$_S$). The output fields of BS$_S$ are written as $a_{S1} = (a_{S_R} + a_{S_L})/\sqrt{2}$ and $a_{S2} = (a_{S_R} - a_{S_L})/\sqrt{2}$, which are directed into detectors D$_{S1}$ and D$_{S2}$, respectively. In the ideal case, a successful detection at detectors D$_{S1}$ or D$_{S2}$ after BS$_S$ projects the two ensembles into an entangled state [48]:

$$\Psi_{R,L}(0) = \left( e^{-i\eta_w} |1\rangle_R |0\rangle_L \pm |0\rangle_R |1\rangle_L \right)/\sqrt{2}, \tag{8}$$

where, $|1\rangle_{R(L)} = \psi_{R(L)}(t=0)$ refers to the SW state in the ensemble *R* (*L*), which corresponds to an excitation, $|0\rangle_{R(L)} = \otimes_j |g_j\rangle_{R(L)}$ denotes the ground state of the ensemble *R* (*L*), $\eta_w = \Delta\phi_w + \Delta\beta_S$, $\Delta\phi_w$ is the phase difference between the write beams from a beam splitter (BS$_w$, which is not marked in **Figure 2**) to the ensembles *R* and *L*, and $\Delta\beta_S$ the phase difference between Stokes (write-out) fields $a_{S_R}$ and $a_{S_L}$ in propagation from the ensembles to the beam splitter BS$_S$. After a storage time *t*, the heralded entanglement state is evolved to

$$\Psi_{R,L}(t) = e^{-i\omega_0 t} e^{-i\delta\varphi_R(t)} D(t) \left( e^{-i\eta_W} |1\rangle_R |0\rangle_L \pm e^{i[\delta\varphi_R(t)-\delta\varphi_L(t)]} |0\rangle_R |1\rangle_L \right) / \sqrt{2}, \qquad (9)$$

where, we assume that Larmor frequencies caused by the bias magnetic field in $R$ and $L$ ensembles have the same value $\omega_0$, the decay factors for $|1\rangle_R$ and $|1\rangle_L$ are assumed to be $D_R(t) = D_L(t) = D(t)$, the retrieval efficiencies $\gamma_R(t) = \gamma_L(t) = \gamma_0 |D(t)|^2 = \gamma(t)$, the phase noise $\delta\varphi_R(t) = 2\pi\mu'\delta B_R t$ and $\delta\varphi_L(t) = 2\pi\mu'\delta B_L t$, $\delta B_R(t)$ and $\delta B_L(t)$ are SMFFs in the ensembles $R$ and $L$, respectively, whose Lorentzian distribution are assumed to have the same rms width $\sigma_R = \sigma_L = \sigma_B$. The entanglement state may be verified following the protocol in Refs [25] and [47], where, read beams are applied into the two ensembles to convert the spin waves $\psi_R(t)$ and $\psi_L(t)$ into anti-Stokes fields $a_{aS_R}$ and $a_{aS_L}$, respectively. In the ideal case, the atomic state $\Psi_{R,L}(t)$ is directly converted to the photonic entangled state $\Phi_{R,L}$. The entanglement degree of $\Psi_{R,L}(t)$ or $\Phi_{R,L}$ can be characterized by the concurrence [25, 47]

$$\mathcal{C} = \max\left(0, \frac{2d - \sqrt{p_{00}p_{11}}}{P}\right), \qquad (10)$$

where, $p_{ij}$ corresponds to the probability to find $i$ photons in the field $a_{aS_L}$ and $j$ photons in the field $a_{aS_R}$, $P = p_{01} + p_{10} + p_{00} + p_{11}$, $d = V(p_{01} + p_{10})/2$ is the coherence term between the states $|1\rangle_R |0\rangle_L$ and $|0\rangle_R |1\rangle_L$, $V$ is the visibility of the interference fringes between the anti-Stokes fields $a_{aS_L}$ and $a_{aS_R}$ when the relative phase $\theta$ between the fields is scanned. The concurrence $\mathcal{C}$ ranges from 0 for a separable state to 1 for a maximally entangled state [25]. The probabilities of $p_{ij}$ can be directly measured via photon counting in the fields $a_{aS_L}$ and $a_{aS_R}$. The measurement of $V$ requires to combine the two retrieved fields on a 50/50 beam splitter, which is labeled as BS$_{aS}$ in Fig. 2. After BS$_{aS}$, the output modes $a_{aS1} = (a_{aS_R} + a_{aS_L})/\sqrt{2}$ and $a_{aS2} = (a_{aS_R} - a_{aS_L})/\sqrt{2}$ are directed to single-photon detectors D$_{aS1}$ and D$_{aS2}$, respectively. The visibility $V$ can be measured as [65]

$$V = \frac{\text{Max}\{P_{S1,\,aS1}(\theta)\} - \text{Min}\{P_{S1,\,aS1}(\theta)\}}{\text{Max}\{P_{S1,\,aS1}(\theta)\} + \text{Min}\{P_{S1,\,aS1}(\theta)\}}, \tag{11}$$

where, $\text{Max}\{P_{S1,\,aS1}(\theta)\}$ ($\text{Min}\{P_{S1,\,aS1}(\theta)\}$) denotes the maximum (minimum) coincidence detection between the Stokes and anti-Stokes fields $a_{S1}$ and $a_{aS1}$ when scanning $\theta$. The probability $P_{S1,\,aS1}(\theta)$ is found to be (see Eq.(S9), Supporting information)

$$P_{S1,\,aS1}(\theta) = \chi \gamma(t) \eta^2 \left(1 + e^{-t/\tau_0} \cos\theta\right)/2 + P_S P_{aS} \tag{12}$$

where, $\chi$ ($\gamma(t)$) is excitation probability (retrieval efficiency) for either ensemble alone, with $\chi_R = \chi_L = \chi$ ($\gamma_R = \gamma_L = \gamma$) assumed, $\eta$ denotes the detection efficiency of the detectors ($D_{S1}$, $D_{aS2}$, $D_{S1}$ and $D_{aS2}$), $\tau_0 = \left[2\pi\mu'\sigma_\Delta\right]^{-1}$ is the lifetime of the coherence between the two SWs, $\sigma_\Delta = 2\sigma_B$ is the relative width of SMFFs between the two ensembles, and $P_S$ ($P_{aS}$) the probability of detecting a photon in the Stokes field $a_S$ (anti-Stokes field $a_{aS}$) for either ensemble, with $P_{S_L} = P_{S_R} = P_S$ ($P_{aS_L} = P_{aS_R} = P_{aS}$) assumed. The visibility is then found to be [see Supporting information for details]

$$V = \frac{g_{S,aS} - 1}{g_{S,aS} + 1} e^{-t/\tau_0}, \tag{13}$$

where, the factor $e^{-t/\tau_0}$ describes exponential decay of the entanglement state with $t$ due to SMFFs, $g_{S,aS}$ denotes the cross-correlation function between Stokes and anti-Stokes fields for either ensemble alone (see Eq.(S11) in Supporting information). In the previous work of Ref. [25], the visibility of the single-excitation entangle state is described as $V = \frac{g_{S,aS} - 1}{g_{S,aS} + 1}$, which doesn't involve the dephasing induced by SMFFs. Contrast to that work, our Eq. (13) shows that the visibility of single-excitation entanglement between two ensembles suffers from not only the decoherence due to atomic motions and the inhomogeneous broadening, but also shot-to-shot phase noise caused by SMFFs.

In the following, we demonstrate a proof-in-principle experiment to verify the inference of Eq. (13).

## 3. Experimentally demonstrating decoherence of single-excitation entanglement between two SW memories caused by SMFFs in a cold atomic ensemble

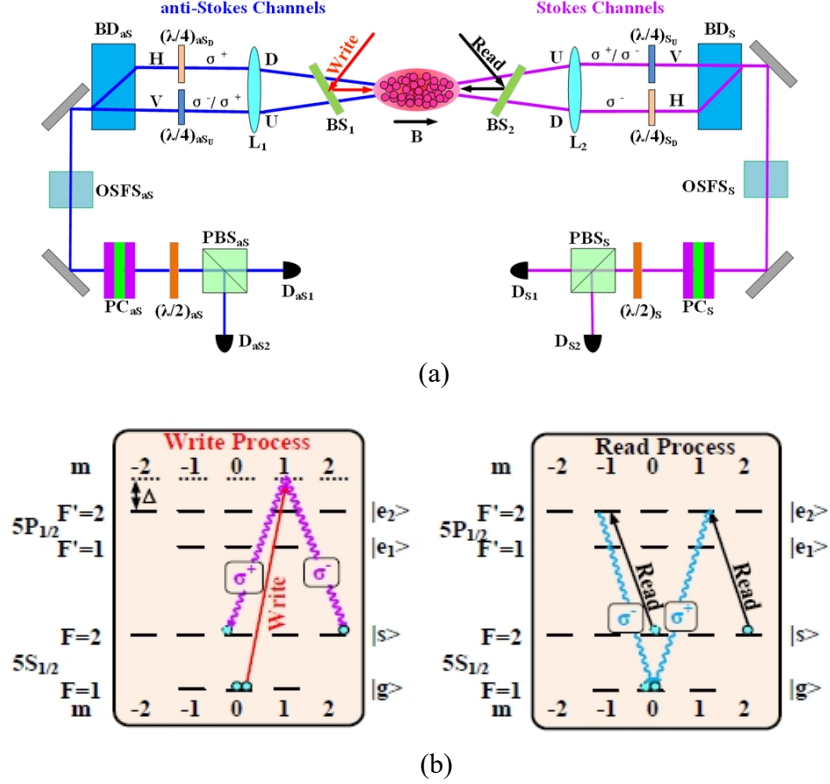

(a)

(b)

**Figure 3.** (a) Experimental setups for generating entanglement states. $BD_S$ ($BD_{aS}$): beam displacer; $BS_1$ ($BS_2$): non-polarizer beam splitter; $L_1$ ($L_2$): lens; $OSFS_S$ ($OSFS_{aS}$): optical spectrum filter systems for Stokes (anti-Stokes) channel; $PC_S$ ($PC_{aS}$): phase compensator [39]; $(\lambda/4)_{S_{U(D)}}$ and $(\lambda/4)_{aS_{U(D)}}$: quarter-wave plates; $(\lambda/2)_{S(aS)}$: half-wave plate. $BD_S$ and $BD_{aS}$ form the Stokes-anti-Stokes interferometer. U and D denote the two arms of the interferometer. A $\sigma^+$-polarized write pulse is applied to the MOT atoms through $BS_1$, which induces spontaneous Raman scattering of $\sigma^+$ and/or $\sigma^-$-polarized Stokes photons and create simultaneously SWs. For the generation of the first (second) entanglement state, the $\sigma^+$ ($\sigma^-$)-polarized photon propagating along U arm is changed into $V$-polarization after $(\lambda/4)_{S_U}$. In the D arm, the $\sigma^-$-polarized photon is changed into $H$-polarization after $(\lambda/4)_{S_D}$. The $V$-polarized photon in U arm and $H$-polarized photon in D arm are combined by $BD_S$. **(b)** Relevant levels for write and read processes.

The experimental setup is illustrated in **Figure 3**.(a), which mainly includes a polarization interferometer and a cloud of cold $^{87}$Rb atoms trapped in a magneto-optical trap (MOT). The interferometer is formed by two beam displacers (BD$_{aS}$ and BD$_S$) with the cold atomic ensemble centered between them. The relevant levels of the atoms are shown in Fig. 3(b), where $|g\rangle=|5S_{1/2},F=1\rangle$, $|s\rangle=|5S_{1/2},F=2\rangle$, $|e_1\rangle=|5P_{1/2},F'=1\rangle$, and $|e_2\rangle=|5P_{1/2},F'=2\rangle$. The experiment is carried out in a cycle fashion. In each cycle, the magneto-optical trap (MOT) is switched on for 23 ms for preparing the cold atoms and then switched off for 10 ms for single-excitation entanglement generation. Before the entanglement generation, we apply a bias magnetic field ($B_0 \approx 1G$) onto the atoms along z-axis by using Helmholtz coils (with a size of ~300mm), which is driven by a DC power supply. At the same time, we apply pump laser pulses to prepare the atoms into initial Zeeman state $|g, m_F=0\rangle$. The magnetic field along z-axis has a slow fluctuation of $\delta B$ with a rms width of $\sigma_B$. We then start entanglement-generation trials. In the beginning of each trial, we apply a $\sigma^+$-polarized write pulse with frequency +140 MHz detuned from $|a\rangle \to |e_2\rangle$ transition through a beam splitter BS$_1$ (see **Figure 3(a)**). The write pulse induces the Raman transition $|g, m_F=0\rangle \to |s, m_F=0\rangle$ ($|g, m_F=0\rangle \to |s, m_F=+2\rangle$) via $|e_2, m_F=+1\rangle$, which emits $\sigma^+$ ($\sigma^-$)-polarized Stokes photons and simultaneously creates SWs associated with the magnetic-field-insensitive (magnetic-field-sensitive) coherence between $|g, m_F=0\rangle \leftrightarrow |s, m_F=0\rangle$ ($|g, m_F=0\rangle \leftrightarrow |s, m_F=+2\rangle$). If a Stokes photon is emitted into the arm $U$ ($D$) of the interferometer, a collective (SW) excitation will be created in the mode $M_U$ ($M_D$) with the wave-vector defined by $k_{M_U}=k_w-k_{S_U}$ ($k_{M_D}=k_w-k_{S_D}$), where, $k_w$ denotes the wave-vector of write beam, $k_{S_U}$ ($k_{S_D}$) denotes the wave-vector of the Stokes photon $S_U$ ($S_D$) emitting into the arm $U$ ($D$). For suppressing atom-motion-induced decoherence [39], we set the angles between the wave-vectors $k_w$ and $k_{S_{U(D)}}$ to be small values of $\pm 0.7^0$.

We generate two kinds of heralded entanglement states between two SWs. The first entangled state stores the two SWs in magnetic-field-insensitive (MFI) and magnetic-field-sensitive (MFS) spin transitions, respectively. We use this state to demonstrate entanglement degradation induced by SMFFs. In this entanglement generation, the $\sigma^+$-polarized ($\sigma^-$-polarized) Stokes photons $S_U$ ($S_D$) go through a wave plate $(\lambda/4)_{S_U}$ ($(\lambda/4)_{S_D}$). By setting the polarization angles of $(\lambda/4)_{S_U}$ and $(\lambda/4)_{S_D}$, we change the $\sigma^+$ and $\sigma^-$-polarized $S_U$ and $S_D$ into $V$ and $H$–polarization, respectively. The $V$–polarized $S_U$ and $H$–polarized $S_D$ fields are combined into a spatial mode $S$ by BD$_S$. The field $S$ goes through a wave plate $(\lambda/2)_S$ and a polarization beam splitter $PBS_S$ (see Fig. 3(a)). With $(\lambda/2)_S$ at $\vartheta = 22.5°$, $PBS_S$ splits the input $S$ field into two modes $a_{S1} = (a_{S_U} + a_{S_D})/\sqrt{2}$ and $a_{S2} = (a_{S_U} - a_{S_D})/\sqrt{2}$, which are directed to detectors $D_{S1}$ and $D_{S2}$, respectively. Upon a detection event at $D_{S1}$ in the ideal case with excitation probability $\chi = \chi_{M_U} \approx \chi_{M_D} \leq 1$, the two SWs are created in $M_U$ and $M_D$ modes and projected into the entangled state:

$$\Psi^{\uparrow,\downarrow}_{M_{U,D}}(0) = \left(e^{-i\Delta\phi_S}|1\rangle^{\uparrow}_{M_U}|0\rangle + |0\rangle|1\rangle^{\downarrow}_{M_D}\right)/\sqrt{2} + O(\chi), \quad (14)$$

where, labels $\uparrow$ and $\downarrow$ denote the SWs associated with the MFI and MFS, respectively, $|1\rangle$ denotes one collective excitation, $\Delta\phi_S$ is the phase difference between two Stokes fields in propagation from the ensemble to BD$_S$. After a storage time $t$, the heralded entanglement state is evolved to:

$$\Psi^{\uparrow,\downarrow}_{M_{U,D}}(t) = D_0(t)e^{-i\omega_U t - i\delta\varphi_U(t)}\left(e^{-i\Delta\phi_S}|1\rangle^{\uparrow}_{M_U}|0\rangle + e^{-i\Delta\omega_0 t - i[\delta\varphi_D(t) - \delta\varphi_U(t)]}|0\rangle|1\rangle^{\downarrow}_{M_D}\right)/\sqrt{2}, \quad (15)$$

where, $\Delta\omega_0 = \omega_{D0} - \omega_{U0}$ denotes the frequency difference of Larmor precessions of SWs between $M_U$ and $M_D$ modes under the bias field $B_0$, $\delta\varphi_{U(D)}(t) = 2\pi\mu'_{U(D)}\delta Bt$ denotes the phase noise of the SW in $M_U$ ($M_D$) mode caused by $\delta B$. The sensitivities of the MFI spin wave $|1\rangle^{\uparrow}_{M_U}$ and MFS spin wave $|1\rangle^{\downarrow}_{M_D}$ can be evaluated as $\mu'_U \approx 0$ and $\mu'_D = \frac{\mu_B}{h}$ respectively. Finally, we have $\delta\varphi_D(t) - \delta\varphi_U(t) \approx 2\pi\frac{\mu_B}{h}\delta Bt$ and $\Delta\omega_0 = \frac{2\pi\mu_B}{h}B_0$. For simplicity, we have assumed that the amplitude-decay factors for $|1\rangle^{\uparrow}_{M_U}$ and $|1\rangle^{\downarrow}_{M_D}$ are the same, i.e. $D^{\uparrow}_{M_U}(t) = D^{\downarrow}_{M_D}(t) = D_M(t)$. The SW entanglement state $\Psi_{M_{U,D}}^{\uparrow,\downarrow}(t)$ can be rewritten as,

$$\Psi_{M_{U,D}}^{\uparrow,\downarrow}(t) = D(t)\left(e^{-i\Delta\phi_S}|1\rangle_{M_U}^{\uparrow}|0\rangle + e^{-i2\pi\mu_B(B_0+\delta B)t/h}|0\rangle|1\rangle_{M_D}^{\downarrow}\right)/\sqrt{2}, \quad (16)$$

By applying a read beam into the ensemble, the spin waves $|1\rangle_{M_U}^{\uparrow}$ and $|1\rangle_{M_D}^{\downarrow}$ will be converted into anti-Stokes fields $a_{aS_U}$ and $a_{aS_D}$. As shown in Fig.3(a), the retrieved field $a_{aS_U}$ from the atoms is $\sigma^-$-polarized, and it is changed into V–polarization after going through a wave plate $(\lambda/4)_{aS_U}$. While, the retrieved field $aS_D$ from the atoms is $\sigma^+$-polarized, and it is changed into H–polarization after a wave plate $(\lambda/4)_{aS_D}$. The anti-Stokes field $a_{aS_U}$ and $a_{aS_D}$ are in a single-photon entangled state:

$$\Phi(t) \propto \left(e^{-i\Delta\phi_S-i\Delta\phi_{aS}}a_{aS_U}^\dagger|0\rangle|0\rangle + e^{-i2\pi\mu_B(B_0+\delta B)t/h}a_{aS_D}^\dagger|0\rangle|0\rangle\right)/\sqrt{2}, \quad (17)$$

where, $\Delta\phi_{aS}$ is the phase difference between $a_{aS_U}$ and $a_{aS_D}$ fields in propagation from the ensemble to $BD_{aS}$. The two anti-Stokes fields are combined into a spatial mode $aS$ after $BD_{aS}$. Then, the mode $aS$ goes through a wave plate $(\lambda/2)_{aS}$ and is directed into a polarization beam splitter $PBS_{aS}$. When measuring the coherence between the two spin waves, $(\lambda/2)_{aS}$ is set at $22.5^0$. At this case, $PBS_{aS}$ splits the input $aS$ field into two modes $a_{aS1}=(a_{aS_U}+a_{aS_D})/\sqrt{2}$ and $a_{aS2}=(a_{aS_U}-a_{aS_D})/\sqrt{2}$, which are directed into the single-photon detectors $D_{aS1}$ and $D_{aS2}$, respectively. After the retrieval, the atoms are prepared in $|g, m_F=0\rangle$ again by the pumping pulses, and then the next trail starts. If no photon is detected by $D_{S1}$, the pumping pulses are directly applied and the next trail starts.

**4. Results**

At first, we measure the cross-correlation function $g_{S,AS}^{+,-}$ ($g_{S,AS}^{-,+}$) between the $\sigma^+$ ($\sigma^-$) -polarized Stokes and $\sigma^-$ ($\sigma^+$) -polarized anti-Stokes photons defined by $g_{S,aS}^{+,-}=P_{S,aS}^{+,-}/(P_S^+ P_{aS}^-)$ ($g_{S,aS}^{-,+}=P_{S,aS}^{-,+}/(P_S^- P_{aS}^+)$), where $P_S^+(P_S^-)$ and $P_{aS}^-(P_{aS}^+)$ are the probabilities of detecting an $\sigma^+$ ($\sigma^-$) -polarized Stokes and $\sigma^-$ ($\sigma^+$) -polarized anti-Stokes photon in U (D) arm, respectively, $P_{S,aS}^{+,-}$ ($P_{S,aS}^{-,+}$) is the coincidence probability between the $\sigma^+$ ($\sigma^-$)-polarized Stokes and $\sigma^-$ ($\sigma^+$)-polarized anti-Stokes photons in U (D) arm. For measuring these quantities, $(\lambda/2)_S$ and $(\lambda/2)_{aS}$ are both set at $0^0$, and the phase difference $\Delta\phi_S$ ($\Delta\phi_{aS}$) is compensated to be zero by using the phase compensator PC$_S$ (PC$_{aS}$). The black square dots in **Figure 4** (a) and (b) are the measured $g_{S,aS}^{+,-}$ ($g_{S,aS}^{-,+}$) data as a function of $t$. One can see that the measured $g_{S,aS}^{+,-}$ data are basically the same as the measured $g_{S,aS}^{-,+}$ data at various $t$ and both decrease with $t$. The red lines in **Figure 4**(a) and 5(b) are the fittings to the measured $g_{S,aS}^{+,-}$ and $g_{S,aS}^{-,+}$ according to (see Eq. (S11) in Supporting information):

$$g_{S,aS}^{+,-} \approx 1+\gamma^{\uparrow}(t)/\left(\chi\gamma^{\uparrow}(t)+\chi\left(1-\gamma^{\uparrow}(t)\right)\xi_{se}+Z_+\right), \quad (18a)$$

$$g_{S,AS}^{-,+} \approx 1+\gamma^{\downarrow}(t)/\left(\chi\gamma^{\downarrow}(t)+\chi\left(1-\gamma^{\downarrow}(t)\right)\xi_{se}+Z_-\right), \quad (18b)$$

where, $\xi_{se}$ denotes the branching ratio corresponding to the read-photon transitions, $Z_\pm$ the background noise in the $\sigma^\pm$-polarized anti-Stokes channels, $\gamma^{\uparrow}$ and $\gamma^{\downarrow}$ denote the retrieval efficiency of the MFI and MFS SWs, respectively,. In the fittings, $\gamma^{\uparrow}$ and $\gamma^{\downarrow}$ values are taken from the measured data in **Fig. S1** in Supporting information.

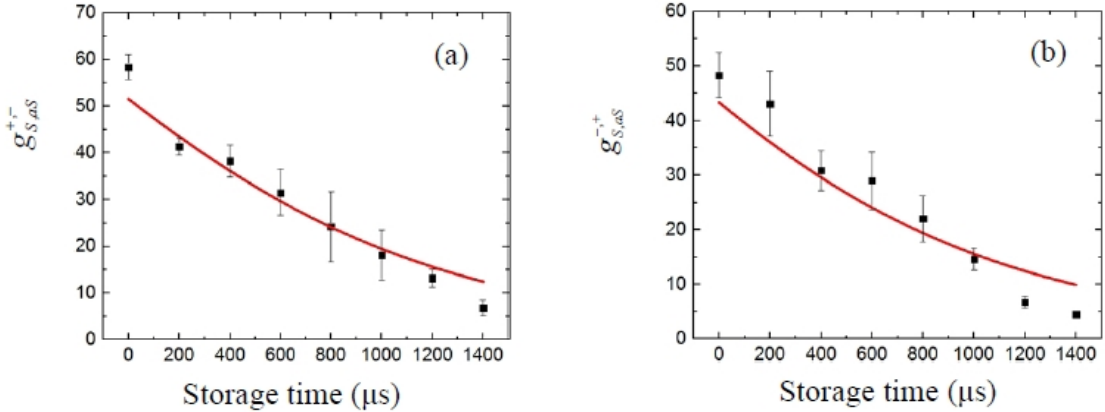

**Figure 4.** (a) and (b) are the measured cross-correlation function $g_{S,AS}^{+,-}$ and $g_{S,AS}^{-,+}$ as a function of storage time $t$, respectively. The red lines are fittings based on Eqs. (18a) and (18b), where, $Z_- = 3.3\times10^{-4}$ and $Z_+ = 3.1\times10^{-4}$ are experimentally measured values. The fittings yield $\zeta_{se} = 0.26$.

To measure the visibility $V^{\uparrow,\downarrow}$ of the entanglement state $\Psi_{M_{U,D}}^{\uparrow,\downarrow}$, we set $(\lambda/2)_S$ and $(\lambda/2)_{aS}$ at $22.5°$. The counts of the retrieved photon at the detector $D_{aS1}$ are recorded conditioned on a detection event at $D_{S1}$. When scanning the relative phase $\theta$ between the two anti-Stokes fields, the counts show an interference fringe and then we obtain the visibility. The black square dots in **Figure 5** are the measured $V^{\uparrow,\downarrow}$ as a function of storage time $t$. The blue star dots in **Figure 5** are the calculated visibility according to $V_g^{\uparrow,\downarrow} = \zeta \frac{\bar{g}_{S,aS}(t)-1}{\bar{g}_{S,aS}(t)+1}$, where, $\bar{g}_{S,aS}(t) = \left(g_{S,aS}^{+-}(t) + g_{S,aS}^{-+}(t)\right)/2$, $\zeta=0.85$ is the overlap between two arms. As shown in **Figure 5**, the measured $V^{\uparrow,\downarrow}$ rapidly decreases and while the calculated $V_g^{\uparrow,\downarrow}$ basically remains unchanged with the increase in the storage time $t$. This deviation is just due to the influence of the SMFFs on the visibility $V^{\uparrow,\downarrow}$. The red solid line is the fitting to the measured $V^{\uparrow,\downarrow}$ according to $V^{\uparrow,\downarrow} = V_g^{\uparrow,\downarrow} \xi' e^{-t/\tau_0}$, where, $\tau_0 : \left(2\pi\mu_B\sigma_B/h\right)^{-1}$ is the lifetime limited to the magnetic-field fluctuation $\delta B$, $\xi'=0.88$ is a non-unit parameter, whose introduce is because that large difference of the Larmor frequency between the two spin waves causes a decrease in the coherence between two retrieval fields. The fittings yield lifetime $\tau_0 \approx 50\mu s$, which corresponds to rms width of $\sigma_B \approx 2.25mG$ according to $\tau_0 \approx \left(2\pi\mu_B\sigma_B/h\right)^{-1}$.

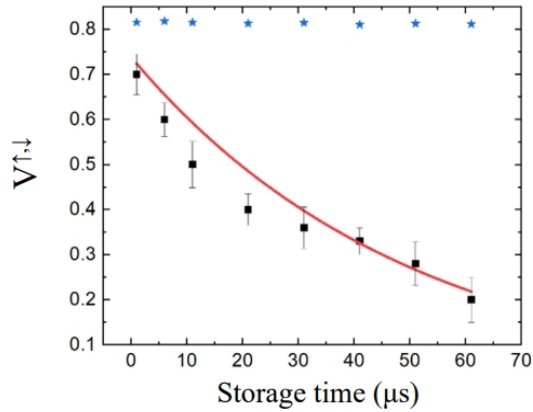

**Figure 5** The measured visibility $V^{\uparrow,\downarrow}$ (black squares) as a fluctuation of storage time $t$. The red line is the fitting according to $V^{\uparrow,\downarrow} = V_g^{\uparrow,\downarrow} \xi' e^{-t/\tau_0}$. For comparison, we plot the calculated visibility

data (blue stars) according to $V_g^{\uparrow,\downarrow} = \zeta \left( \bar{g}_{S,aS}(t) - 1 \right) / \left( \bar{g}_{S,AS}(t) + 1 \right)$.

To demonstrate a scheme in which the phase difference noises between two SWs can be eliminated, we generate the second entanglement state that stores two SW modes in the MFS transitions. For which, we reset $(\lambda/4)_{S_U}$ in U arm, which makes the $\sigma^-$-polarized $S_U$ photon change into $V$-polarization. The $V$-polarized $S_U$ field and the $H$-polarized $S_D$ field combine into $S$ mode after BD$_S$. Conditioned on a detection event at $D_{S1}$, the second entanglement state $\Psi_{M_{U,D}}^{\downarrow,\downarrow}(0) = \frac{1}{\sqrt{2}} \left( e^{-i\phi_s} |1\rangle_{M_U}^{\downarrow} |0\rangle + |0\rangle |1\rangle_{M_D}^{\downarrow} \right) + O(\chi)$ is created, where, $|1\rangle_{MU}^{\downarrow}$ ($|1\rangle_{MD}^{\downarrow}$) represents one SW excitation associated with the MFS coherence in $M_{U\,(D)}$ mode. After a storage time $t$, the heralded entanglement state $\Psi_{M_{U,D}}^{\downarrow,\downarrow}(0)$ is evolved into:

$$\Psi_{M_{U,D}}^{\downarrow,\downarrow}(t) = D(t)e^{-i\omega_0 t}e^{-i\delta\varphi(t)} \left( e^{-i\Delta\phi_s} |1\rangle_{M_U}^{\downarrow} |0\rangle + |0\rangle |1\rangle_{M_D}^{\downarrow} \right) / \sqrt{2} , \tag{19}$$

where, $\omega_0$ denotes the Larmor frequency caused by the bias magnetic field $B_0$ for either spin-wave modes alone with $\omega_0 = \omega_{U_0} = \omega_{D_0} = 2\pi \frac{\mu_B}{h} B_0$, the phase noise caused by SMFFs for mode M$_U$ and M$_D$ is the same i.e., $\delta\varphi(t)_{M_U} = \delta\varphi(t)_{U_D} = \delta\varphi(t) = 2\pi\mu'\delta B t$. Thus, the visibility of the entanglement state $\Psi_{M_{U,D}}^{\downarrow,\downarrow}$ in Eq. (19) is expressed as

$$V^{\downarrow,\downarrow} = \zeta \frac{g_{S,aS}^{-,+} - 1}{g_{S,aS}^{-,+} + 1} = V_g^{\downarrow,\downarrow} , \tag{20}$$

which is not affected by SMFFs. Similar to the retrieval of the entangled state $\Psi_{M_{U,D}}^{\uparrow,\downarrow}(t)$, we apply the read beam to map the SWs $|1\rangle_{M_U}^{\downarrow}$ and $|1\rangle_{M_D}^{\downarrow}$ to the anti-Stokes photon in U and D arms, respectively. In the retrieval of $\Psi_{M_{U,D}}^{\downarrow,\downarrow}(t)$, the retrieved field in U arm is $\sigma^+$-polarized. So, we reset the wave plate $(\lambda/4)_{aS_D}$, which changes $\sigma^-$-polarized anti-Stokes photon into V-polarization. The $H$-polarized anti-Stokes field in D arm and $V$-polarized anti-Stokes field in U arm are combined into a$S$ mode by BD$_{aS}$. Similar to the measurement for $V^{\uparrow,\downarrow}$, we measure the visibility $V^{\downarrow,\downarrow}$ of the entanglement state $\Psi_{M_{U,D}}^{\downarrow,\downarrow}$. The black squares in **Figure 6** are the measured $V^{\downarrow,\downarrow}$ as a function of storage time $t$. The red circles are the calculated visibilities according to $V^{\downarrow,\downarrow} = \zeta \frac{g_{S,AS}^{-,+}(t)-1}{g_{S,AS}^{-,+}(t)+1}$, where $\zeta = 0.84$. One can see that the measured data are well in agreement with the calculated data for various $t$, showing that the phase difference noise between the two SWs is eliminated.

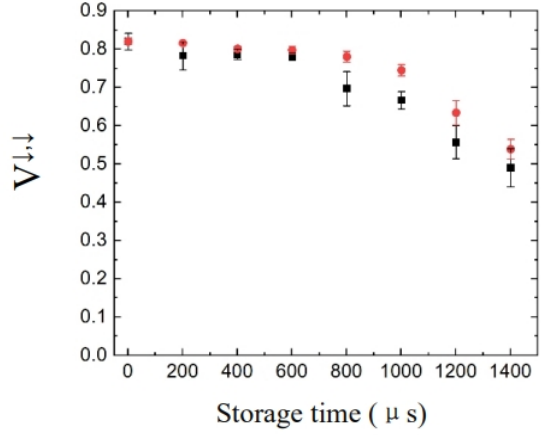

**Figure 6.** The measured visibilities $V^{\downarrow,\downarrow}$ (black squares) and the calculated visibilities (red circles) according to $V^{\downarrow,\downarrow} = \zeta \left(g_{S,AS}^{-,+}(t)-1\right)/\left(g_{S,AS}^{-,+}(t)+1\right)$ as a function of the storage time $t$.

As pointed out in the above, the entanglement degree of the states $\Psi_{M_{U,D}}^{\uparrow,\downarrow}(t)$ and $\Psi_{M_{U,D}}^{\uparrow,\downarrow}(t)$ are characterized by concurrence $C = \max(0, V(p_{01} + p_{10}) - \sqrt{p_{00}p_{11}}/P)$, where, $P = p_{01} + p_{10} + p_{00} + p_{11}$, $p_{ij}$ corresponds to the probability to detect $i$ anti-Stokes photons in U mode and $j$ anti-Stokes photons in D mode. For the case $\chi \ll 1$, $p_{10} = p_{01} = p_c/2$, where, $p_c \approx \gamma(t)\eta$ is the conditional probability of detecting a photon in the anti-Stokes field from one spin-wave mode following a detection event for Stokes field, $\eta$ is the efficiency for each detection channel.

The black squares in **Figure 7**(a) and **7**(b) are the measured concurrences $C^{\uparrow,\downarrow}$ and $C^{\downarrow,\downarrow}$ for the states $\Psi_{M_{U,D}}^{\uparrow,\downarrow}(t)$ and $\Psi_{M_{U,D}}^{\downarrow,\downarrow}(t)$, respectively, as a function of the storage time $t$. The red and blue lines in **Figure 7**(a) and **7**(b) are the calculated data according to Ref. [25]

$$C^{\uparrow,\downarrow} \simeq \max\left[0, \overline{p_c}(V - 2\sqrt{(1-\overline{p_c})/\overline{g}_{S,aS}})\right], \quad (21a)$$

$$\mathcal{C}^{\downarrow,\downarrow} \simeq \max\left[0, p_c(V - 2\sqrt{(1-p_c)/g_{S,aS}^{-,+}})\right], \quad (21b)$$

respectively, where, $\bar{p}_c = \eta(\gamma^\uparrow + \gamma^\downarrow)/2$, $\bar{g}_{S,aS} = (g_{S,aS}^{+,-} + g_{S,aS}^{-,+})/2$, $\gamma^\uparrow$ ($\gamma^\downarrow$) denotes the retrieval efficiency from the MFI (MFS) SW. The fittings are in agreement with the measured data. The concurrences for both entanglement states degrade with $t$. One reason for this is the SW decoherence caused by the atomic random motions and magnetic-field gradient, which lead to decays of the cross-correlation functions $g_{S,aS}$ with $t$ (**Figure 4**). The storage time for the concurrence $\mathcal{C}^{\uparrow,\downarrow} \geq 0$ is ~40 us, which is far less than that for the concurrence $\mathcal{C}^{\downarrow,\downarrow} \geq 0$ (~1.2 ms). We attributed this to that the entanglement state $\Psi_{M_{U,D}}^{\uparrow,\downarrow}(t)$ suffers from dephasing caused by SMFFs while the entanglement state $\Psi_{M_{U,D}}^{\downarrow,\downarrow}(t)$ does not.

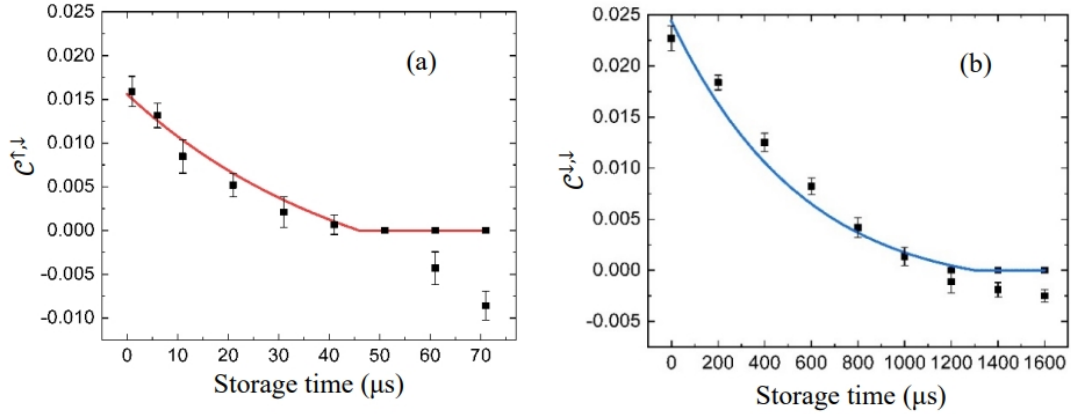

**Figure 7.** The black squares are the measured concurrences $\mathcal{C}^{\uparrow,\downarrow}$ (a) and $\mathcal{C}^{\downarrow,\downarrow}$ (b) as a function of storage time $t$. The red and blue lines are obtained from Eqs. (21a) and (21b), the retrieval efficiencies data come from Fig. S1.

## 5. Evaluating decoherence rate of single-excitation entanglement between two ensembles for various width of SMFFs

The protocol of the entanglement establishment between two ensembles has been shown in **Figure 2**. We assume that each repeater node (L or R ensemble) uses the state-of-art atomic SW memory system, where the atoms are loaded in an optical lattice and coupled into a cavity. In this case, the node memory has a single-mode initial efficiency of $\gamma_0 \approx 76\%$ and a lifetime of $\tau_D \approx 410$ ms [38]. Its time-dependent retrieval efficiency can be approximately written as $\gamma(t) = \gamma_0 e^{-t/\tau_D}$. When obtaining such a long-lived SW memory, a bias magnetic field (~4G) has to be applied onto the atoms to compensate the differential light shift [38]. The bias magnetic field can be produced by using Helmholtz coils driven with a DC power supply, which also generate SMFFs. In the individual ensembles, the rms width of SMFFs may be reduced to $\sigma_B \approx 2 \text{ mG}$ by using low-noise supply, which is demonstrated in our presented experiment. It may also be reduced to $\sigma_B \approx 1 \text{ mG}$ by using an active magnetic-field stabilization in Ref. [63] and $\sigma_B \approx 0.2 \text{ mG}$ via compensating the recorded magnetic-field noise with an open-loop feed-forward circuit in Ref. [64]. In common cases, two current supplies are individually applied to R and L Helmholtz coils, respectively, and then SMFFs in R and L nodes are independent. So, the relative width of SMFFs between the two ensembles is $\sigma_\Delta = 2\sigma_B$. The sensitivity of the used clock coherence to the magnetic field in optical lattice atoms has been measured in Ref.[62], which is $\mu' \approx 5 \text{Hz/mG}$. Using the above parameters, we calculate the dependence of concurrence of entanglement for a repeater link on storage time $t$, which is shown in **Figure 8**. The green, blue, and red solid curves are the calculated $\mathcal{C}$ vs $t$ for the magnetic-fluctuation differences with widths $\sigma_\Delta = 4 \text{ mG}$, $\sigma_\Delta = 2 \text{ mG}$, and $\sigma_\Delta = 0.4 \text{ mG}$. From these curves, we extract entanglement lifetimes (the storage time for $\mathcal{C} > 0$) and show them in Table I. As shown in Table I, the lifetimes increase with the decrease in the widths, which shows the influence of SMFFs on the single-excitation entanglement. We propose a scheme to overcome such influence on the single-excitation entanglement storage in one link. Contrast to the common schemes, our scheme applies a single current supply to the Helmholtz coils at $R$ and $L$ nodes, which are connected in series. At each node, the slow magnetic-field fluctuation $\delta B(t)$ is proportional to the current fluctuations $\delta I(t)$ of the Helmholtz coils, i.e., $\delta B(t) \propto \delta I(t)$. We assume that the link has a typical length (separation distance between $R$ and $L$ nodes) of $d = 50 \text{ km}$. In this case, the frequency (with a range of 200-0Hz) of SMFFs is far less than $\frac{c}{d} \sim 6 \text{ kHz}$, which means that the current is quasi-steady, i.e., $\delta I_R(t) \approx \delta I_L(t)$. Also, we assumed that the scheme uses $\mu$-metals to shield the residual field in $R$ and $L$ ensembles, respectively. In these cases, the difference between magnetic-field fluctuations in R and L nodes may be effectively eliminated, meaning that the width difference $\sigma_\Delta \to 0$. The black line in **Figure 8** is the calculated concurrence of the single-excitation entanglement as storage time $t$ for $\sigma_\Delta \to 0$, one can see that the lifetime of entanglement storage ($\mathcal{C}>0$) reaches to 1.7 s.

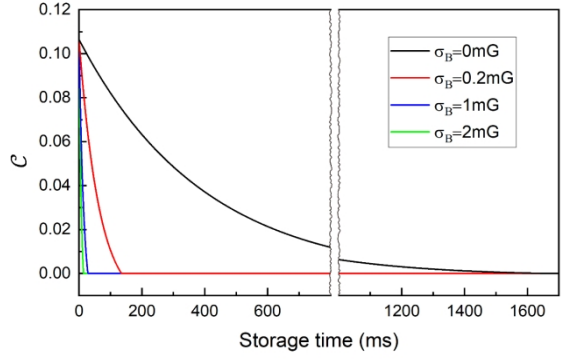

**Figure 8**. The calculated concurrence of single-excitation entanglement for a repeater link as a function of storage time $t$ according to $\mathcal{C} = \max[0, p_c(V - 2\sqrt{(1-p_c)/g_{S,aS}})]$, where, the visibility $V = \zeta(g_{S,aS} - 1)/(g_{S,aS} + 1)e^{-t/\tau_0}$, $\tau_0 = [2\pi\mu'(2\sigma_B)]^{-1} = (2\pi\mu'\sigma_\Delta)^{-1}$ is the lifetime limited by SFMMs in individual ensembles, $g_{S,aS} = 1 + \gamma(t)/(\chi\gamma(t) + \chi(1-\gamma(t))\xi_{se} + Z)$ is the cross-correlation function for either ensemble alone, $\zeta = 0.85$, the conditional probability $p_c = \gamma(t)\eta$, the retrieval efficiency is $\gamma(t) = \gamma_0 e^{-t/\tau_D}$ with $\gamma_0 = 76\%$ and $\tau_D = 410$ ms, the excitation probability and the background noise are assumed to be $\chi = 0.5\%$ and $Z = 3 \times 10^{-4}$ per write pulse.

The quantum link efficiency $\eta_{link}$ is defined as $\eta_{link} = T_S/T_G$, where $T_G$ is the lifetime of entanglement storage for a link, $T_S$ is the storage time of entanglement ($\mathcal{C} > 0$). $\eta_{link} \geq 1$ promises a deterministic entanglement delivery [17], which is critical for QR scaling. For a 50-km link, the required time for generating entanglement over the link via SPI is evaluated to be $T_G \approx 630\ ms$ [38]. We calculate the lifetimes of single-excitation entanglement and the quantum link efficiencies for different $\sigma_\Delta$ and list them in **Table 1.** One can see that for the case of $\sigma_\Delta \to 0$, $\eta_{link} \approx 2.70$, which is greater than the threshold ($\eta_{link} \approx 1$) in Ref. [17].

**Table 1** The lifetimes of storing single-excitation entanglement $T_S$ and quantum link efficiency $\eta_{link}$ for various $\sigma_\Delta$

| $\sigma_B$ | 2 mG | 1 mG | 0.2 mG | 0 mG |
|---|---|---|---|---|

| $\sigma_A = 2\sigma_B$ | 4 mG | 2 mG | 0.4 mG | 0 mG |
|---|---|---|---|---|
| $T_S$ | ~14 ms | ~28 ms | ~135 ms | ~1700 ms |
| $\eta_{link} = T_S / T_G$ | ~0.02 | ~0.04 | ~0.21 | ~2.70 |

**6. Discussion and conclusion**

We theoretically develop a SW model which includes not only the amplitude decoherence caused by atomic motions and inhomogeneous broadening of spin transition, but also phase noise induced by SMFFs. For a single-excitation entanglement between two atomic ensembles which forms a repeater link, we identify that the SMFF-induced phase noise will lead to entanglement decoherence. Such decoherence mechanism has not been pointed out in previous works on QRs [2, 22-24]. With a cold atomic ensemble, we experimentally produce a single-excitation entanglement state between two SWs via DLCZ scheme. We then demonstrate that SMFFs induce entanglement decoherence. For a repeater link which use optical-lattice atoms as nodes, we evaluate the lifetimes of storing single-excitation entanglement. The evaluated entanglement storage lifetime (~135 ms) is mainly limited to the SMFF-induced decoherence, even if the SMFF width is reduced to a very small value ($\sigma_B = 0.2 \text{ mG}$) by using a state-of-art noise-compensation technique [64]. We noted that the fluctuation width can be further reduced (for example, up to $\sigma_B = 0.1 \text{ mG}$) via synchronizing the experimental sequence with the mains electricity frequency 50-Hz in that experiment [56,64]. However, when using such noise-suppression technology in entanglement generation over a link, the experimental repeater rate has to be set at 50-Hz, which corresponds to a period of 20 ms. This period is far longer than the communication time ($L / c$) in a typical link, e.g., for a link with length $L$=50 km, $L/c$=0.25 ms. So, the use of such technology will greatly decrease the entanglement generation rate in the link. To eliminate SMFF-induced entanglement decoherence, we propose an approach, in which, the two Helmholtz coils used for the nodes in a link are connected in series and driven by a single DC supply. Such arrangements will eliminate relative phase noise between two node memories and extend the lifetime of entanglement storage to that of SW memories.

An alternative effective method that protect the single-excitation entanglement from decoherence caused by SMFFs [12,17] is dynamical decoupling (DD) strategy. DD strategy realizes the protection by applying pulse sequence to repeatedly rotate two states of memory. Using this strategy, the storage lifetime of single-excitation entanglement between two remote single quantum systems (NV centers) has been significantly extended [17]. Recently, the storages of photonic time-bin qubits in a solid-state ensemble [66] and that of a weak laser pulse in cold atoms [62] used DD strategy to suppress inhomogeneous spin dephasing and then extend memory lifetime. The application of DD pulses to the ensembles will also suppress the SW phase noise caused by SMFFs. However, in atomic-ensemble-based memories, the DD application will lead to SW errors and introduce noise at memory output [66]. So, when using DD sequence including multiple pulses to suppress single-excitation entanglement decoherence in ensembles, the introduced noise will degrade the entanglement degree.

In summary, our work presents effective methods to suppress single-excitation entanglement decoherence induced by SMFFs in atomic-ensemble-based repeater link. Also, our results would be of benefit in the optimal entanglement distribution policies in repeater chains [67].

**Supporting Information**

Supporting Information is available from the Wiley Online Library or from the author.

**Acknowledgements**

This work is supported by the National Natural Science Foundation of China (12174235), the Fund for Shanxi Key Subjects Construction (1331), and the Fundamental Research Program of Shanxi Province (202203021221011)

**Competing Interest**

The authors declare no competing interests.

# Supporting Information

*Can Sun, Ya Li, Yi-bo Hou, Min-jie Wang, Shu jing Li\* , Hai Wang\**

*D. Sun, Y. Li, Y. Hou, M. Wang, S. Li. and H. Wang*

*2. State Key Laboratory of Quantum Optics and Quantum Optics Devices, Institute of Opto-Electronics. 2.Collaborative Innovation Center of Extreme Optics, Shanxi University, Taiyuan 030006, China*

*\*Corresponding author email: lishujing@sxu.edu.cn and wanghai@sxu.edu.cn*

**The dependence of SW amplitude factor on the storage time**

In the main text, SW amplitude factor is defined as $D(t) = \sum_{j=1}^{N} e^{-i\Delta\omega(z_j)t - i\Delta k v_j t}$ with $\Delta\omega(z_j) = 2\pi\mu' B' z_j$. Considering that the atomic velocities follows Boltzmann distribution $f(v_z) = \exp(-mv_z^2 / 2k_B T)$, and the density of atomic cloud follows Gaussian distribution $f(z) = \exp(-z^2 / 2l)$, where, $T$ is the temperature of the atoms, $l$ is the length of the atomic cloud, we have

$$D(t) = \int_{-\infty}^{\infty} f(z) e^{-i2\pi\mu' B' z t} dz \int_{-\infty}^{\infty} f(v) e^{-i\Delta k v t} dv = e^{-t^2/2\tau_1^2} e^{-t^2/2\tau_2^2}, \quad (S1)$$

where, $\tau_1 \approx 1/\Delta k v_s$ [23] and $\tau_2 \approx (2\pi\mu' B' l)^{-1}$ [22], with $v_s$ being the average speed of the atoms.

**The probability of detecting a Stokes-anti-Stokes coincidence $P_{S1,aS1}(\theta)$ between the detectors $D_{S1}$ and $D_{aS1}$**

As shown in Fig. 2 in the main text, the Stokes fields $a_{S_L}$ and $a_{S_R}$ emitted from the *L* and *R* ensembles, respectively, are combined at the beam splitter $BS_S$. After $BS_S$, the output two modes $a_{S1} = (a_{S_R} + a_{S_L})/\sqrt{2}$ and $a_{S2} = (a_{S_R} - a_{S_L})/\sqrt{2}$ are directed into the detectors $D_{S1}$ and $D_{S2}$, respectively. Conditioned on the detection of a Stokes photon at the detector $D_{S1}$, the two ensembles are projected into the single-excitation entanglement state $\Psi_{R,L}(0)=\left(e^{-i\eta_w}|1\rangle_R|0\rangle_L\pm|0\rangle_R|1\rangle_L\right)/\sqrt{2}$, which has been described by Eq. (8) in the main text.

After a storage time $t$, the heralded entanglement state is evolved to

$$\Psi_{R,L}(t)=D(t)\left(e^{-i\eta_w}|1\rangle_R|0\rangle_L\pm e^{i[\delta\varphi_R(t)-\delta\varphi_L(t)]}|0\rangle_R|1\rangle_L\right)/\sqrt{2}, \qquad (S2)$$

which corresponds to the Eq. (9) in the main text. By applying two read laser beams onto $R$ and $L$ ensembles, respectively, the collective excitation $|1\rangle_R$ or $|1\rangle_L$ is converted into an anti-Stokes photon in the mode $a_{aS_R}$ or $a_{aS_L}$, whose propagation paths are shown in Fig. 2 in the main text. So, the atomic entanglement state is mapped into a single-photon entanglement state. The two anti-Stokes modes $a_{aS_R}$ and $a_{aS_L}$ are combined at the beam splitter BS$_{aS}$. Before BS$_{aS}$, the single-photon entanglement state is written as:

$$\Phi_{R,L}(t)\propto\sqrt{\gamma(t)}\left(e^{-i\eta_w-i\eta_r}a_{aS_R}^\dagger|0\rangle|0\rangle\pm e^{i[\delta\varphi_R(t)-\delta\varphi_L(t)]}a_{aS_L}^\dagger|0\rangle|0\rangle\right)/\sqrt{2}, \qquad (S3)$$

where, $\gamma(t)=\gamma_0|D(t)|^2$ is the retrieval efficiency, $\eta_r=\Delta\phi_r+\Delta\beta_{aS}$, $\Delta\phi_r$ is the phase difference between the read beams from a beam splitter (BSr, which is not marked in Fig. 2) to the ensembles R and L, and $\Delta\beta_{aS}$ is the phase difference between read-out fields $a_{aS_R}$ and $a_{aS_L}$ in propagation from the ensemble to the beam splitter BS$_{aS}$. In experiment, the total relative phase $\eta_w+\eta_r$ is actively stabilized to be $2n\pi$, with $n$ being an integer. For which, one needs to stabilize the length differences between the two arms in Stokes-anti-Stokes and write-read-beam interferometers, where, the Stokes-anti-Stokes (write-read-beam) interferometer is formed by beam splitters BS$_S$ and BS$_{aS}$ (BS$_w$ and BS$_r$), as shown in Fig. 2.

After BS$_{aS}$, the two anti-Stokes output modes $a_{aS1}=\left(a_{aS_R}+a_{aS_L}\right)/\sqrt{2}$ and $a_{aS2}=\left(a_{aS_R}-a_{aS_L}\right)/\sqrt{2}$ are directed to the detectors D$_{aS1}$ and D$_{aS2}$, respectively. We rewrite the single-photon entanglement state as:

$$\Phi(t)=\frac{\sqrt{\gamma(t)}}{2}\left(a_{aS1}^\dagger(1+e^{-i(\delta\varphi(t)+\theta)})+a_{aS2}^\dagger(1-e^{i(\delta\varphi(t)+\theta)})\right)|0\rangle, \qquad [S4]$$

where, $\theta$ is the adjustable phase difference between the two modes $a_{aS_R}$ and $a_{aS_L}$, and can be obtained by using a polarization phase compensator, $\delta\varphi(t)=\delta\varphi_L(t)-\delta\varphi_R(t)=2\pi\mu'(\delta B_L - \delta B_R)t$ in which the SMFFs in R and L nodes are assumed to be independent. The conditional probability of detecting a photon in anti-Stokes field $a_{aS1}$ following a detection event for Stokes field $a_{S1}$ can be expressed as

$$P_c(\theta) = \eta\,\gamma(t)\left\langle\left|1+e^{-i(\Delta\varphi(t)+\theta)}\right|^2\right\rangle/4, \quad [S5]$$

where, $\eta$ is the efficiency for each detection channel, and

$$\begin{aligned}&\left\langle\left|1+e^{-i(\Delta\varphi(t)+\theta)}\right|^2\right\rangle/4\\&=\int_{-\infty}^{+\infty}\int_{-\infty}^{+\infty}P(\delta B_L)P(\delta B_R)\left|1+e^{-i(\Delta\varphi(t)+\theta)}\right|^2/4\,d(\delta B_L)d(\delta B_R)\\&=\int_{-\infty}^{+\infty}\int_{-\infty}^{+\infty}P(\delta B_L)P(\delta B_R)\left[2+e^{-i(\Delta\varphi(t)+\theta)}+e^{i(\Delta\varphi(t)+\theta)}\right]/4\,d(\delta B_L)d(\delta B_R)\\&=(1+\cos\theta\,e^{-(t/\tau_0)})/2\end{aligned} \quad [S6]$$

where, $P(\delta B_{L(R)})=\dfrac{\sigma_B/\pi}{\delta B_{L(R)}^2+\sigma_B^2}$ is the magnetic-field fluctuations $\delta B_R$ and $\delta B_L$, $\sigma_B$ is the rms width of the Lorentzian distribution, $\tau_0=(4\pi\mu'\sigma_B)^{-1}$ is the lifetime of the coherence between the two SWs due to SMFFs.

The unconditional probability of detecting a photon in the Stokes field $a_S$ and the anti-Stokes field $a_{aS}$ for either ensemble may be written as

$$P_S = \chi\eta, \quad (S7a)$$

$$P_{aS} = \chi\gamma(t)\eta + \chi(1-\gamma(t))\xi_{se}\eta + Z\eta, \quad (S7b)$$

where, $\eta$ is the detection efficiency for either channel alone, the second term in (S7b) denotes the noise due to the imperfect retrieval efficiency [Lukas Heller, Pau Farrera, Georg Heinze, and Hugues de Riedmatten, Phys. Rev. Lett. 124, 210504 (2020)], $\xi_{se}$ the branching ratio corresponding to the read-photon transitions, $Z$ the background noise in either anti-Stokes channel alone. The unconditional probability of detecting a photon in the

Stokes field $a_{S1}$ and the anti-Stokes field $a_{aS1}$ are expressed as

$$P_{S1} = \frac{1}{2} \times (2\chi\eta), \quad [S8a]$$

$$P_{aS1} = \frac{1}{2} \times \left[ 2\left( \chi\gamma(t)\eta + \chi(1-\gamma(t))\xi_{se}\eta + Z\eta \right) \right], \quad [S8b]$$

where, the factor 1/2 correspond to the 50% chance that the photon is reflected or transmitted at the beam splitter ($BS_S$ or $BS_{aS}$), while the factor 2 result from the symmetry of the scheme where the photon can come from either ensemble. The probability of detecting a coincidence between Stokes and anti-Stokes fields $a_{S1}$ and $a_{aS1}$ can be written as

$$P_{S1,aS1}(\theta) = P_{S1}P_c(\theta) + P_{S1}P_{aS1}$$
$$= \chi\gamma(t)\eta^2 \left(1 + e^{-t/\tau_0}\cos\theta\right)/2 + \chi^2\gamma(t)\eta^2 + \chi^2(1-\gamma(t))\xi_{se}\eta^2 + \chi Z\eta^2, \quad [S9]$$

Introducing the coincidence probability $P_{S1,aS1}(\theta)$ into the Eq. (11) in the main text, we obtain

$$V = \frac{\gamma(t)/2}{\gamma(t)/2 + \chi\gamma(t) + \chi(1-\gamma(t))\xi_{se} + Z} e^{-t/\tau_0} = \frac{g_{S,aS}-1}{g_{S,aS}+1} e^{-t/\tau_0} \quad [S10]$$

where,

$$g_{S,aS} = P_{S,aS}/(P_S P_{aS}) \approx 1 + \frac{\gamma(t)}{\chi\gamma(t) + \chi[1-\gamma(t)]\xi_{es} + Z}, \quad [S11]$$

denotes the cross-correlation function between Stokes and anti-Stokes fields for either ensemble alone, and $P_{S,aS} = \chi\gamma(t)\eta^2 + P_{aS}P_S$ denotes the probability of detecting a coincidence between the Stokes and anti-Stokes fields for either ensemble alone.

**The measured retrieval efficiencies of the two spin waves**

The retrieval efficiencies of MFI and MFS spin waves are measured as $\gamma^{\uparrow} \approx P_{S,aS}^{+-}/(\eta P_S^+)$ and $\gamma^{\downarrow} \approx P_{S,aS}^{-+}/(\eta P_S^-)$, where, $P_{S,aS}^{+-}$ ($P_{S,aS}^{-+}$) is the coincidence probability between the $\sigma^+$ ($\sigma^-$) -polarized Stokes and $\sigma^-$ ($\sigma^+$) -polarized anti-Stokes photons in U (D) arm. Noted that the anti-Stokes photon retrieved from MFI (MFS) spin wave is $\sigma^-$ ($\sigma^+$) -polarized, which is paired with the $\sigma^+$ ($\sigma^-$)-polarized Stokes photon. In the measurements, the polarization angles of $(\lambda/2)_S$ and $(\lambda/2)_{aS}$ are all set at $0^0$, which make the H- and V- polarized Stokes(anti-Stokes) photon before $(\lambda/2)_S$ ($(\lambda/2)_{aS}$) direct into $D_{S1}$ ($D_{aS1}$) and $D_{S2}$ ($D_{aS2}$). The black squares in Fig. S1(a) and S1(b) are the measured $\gamma^\uparrow$ and $\gamma^\downarrow$ as a function of storage time $t$, respectively. The solid lines are the fittings to the measured data according to $\gamma^\uparrow(t) = \gamma_0^\uparrow e^{-t/\tau_D}$ and $\gamma^\downarrow = \gamma_0^\downarrow e^{-t/\tau_D}$ with $\gamma_0^\uparrow = 22\%$, $\gamma_0^\downarrow = 17\%$ and a lifetime of $\tau_D = 1ms$.

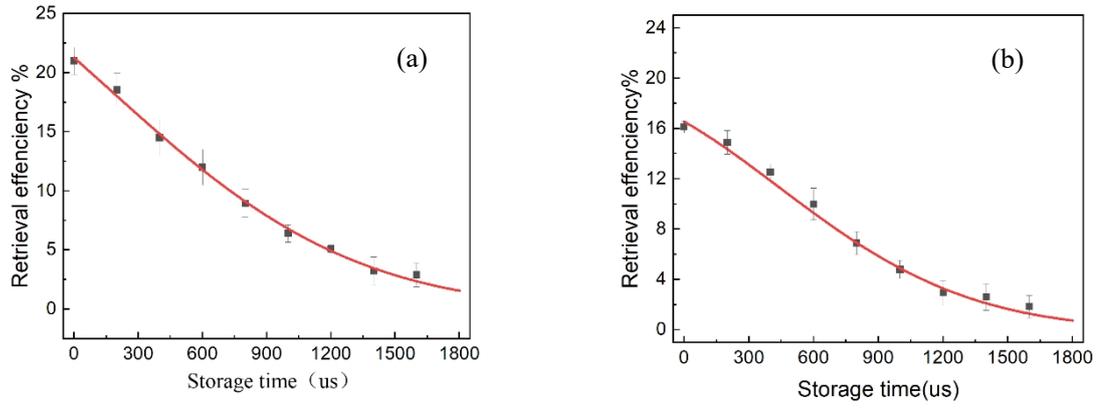

Fig. S1 (a) and S1 (b) are the measured $\gamma^\uparrow$ and $\gamma^\downarrow$ as a function of storage time $t$.